\documentclass[aps,twocolumn,showpacs,10pt]{revtex4-1}
\usepackage[colorlinks,citecolor=blue,linkcolor=red,dvipdfm]{hyperref}
\usepackage{amsmath,amssymb,graphicx}
\usepackage{times}

\begin{document}

\title{Suppression of the critical collapse for one-dimensional solitons by
saturable quintic nonlinear lattices}
\author{Jincheng Shi$^{1,2,3}$}
\author{Jianhua Zeng$^{1,2}$}
\email{\underline{zengjh@opt.ac.cn}}
\author{Boris A. Malomed$^{4,5}$}
\email{\underline{malomed@post.tau.ac.il}}
\affiliation{$^{1}$State Key Laboratory of Transient Optics and Photonics, Xi'an
Institute of Optics and Precision Mechanics of CAS, Xi'an 710119, China}
\affiliation{$^{2}$University of Chinese Academy of Sciences, Beijing 100084,
China}
\affiliation{$^{3}$Key Laboratory for Physical Electronics and Devices of the Ministry of Education \&
Shaanxi Key Lab of Information Photonic Technique, Xi'an Jiaotong University, Xi'an 710049, China}
\affiliation{$^{4}$Department of Physical Electronics, School of Electrical
Engineering, Faculty of Engineering, Tel Aviv University, Tel Aviv 69978,
Israel} \affiliation{$^{5}$ ITMO University, St. Petersburg 197101, Russia}

\begin{abstract}
The stabilization of one-dimensional solitons by a nonlinear lattice against
the critical collapse in the focusing quintic medium is a challenging issue.
We demonstrate that this purpose can be achieved by combining a nonlinear
lattice and saturation of the quintic nonlinearity. The system supports
three species of solitons, namely, fundamental (even-parity) ones and dipole
(odd-parity) modes of on- and off-site-centered types. Very narrow
fundamental solitons are found in an approximate analytical form, and
systematic results for very broad unstable and moderately broad partly
stable solitons, including their existence and stability areas, are produced
by means of numerical methods. Stability regions of the solitons are
identified by means of systematic simulations. The stability of all the
soliton species obeys the Vakhitov-Kolokolov criterion.
\end{abstract}

\maketitle


\textbf{The critical collapse is one of fundamental aspects of the
nonlinear-wave theory. Manifestations of the collapse play a significant
role in various areas of physics, especially in nonlinear optics and studies
of matter waves in atomic Bose-Einstein condensates. The occurrence of the
critical collapse means that the configuration, governed by the nonlinear
Schr\"{o}dinger equation (NLSE)\ in dimension }$D$\textbf{, with power }$p$%
\textbf{\ of the nonlinear term, develops a singularity after a finite
evolution time, if }$D$\textbf{\ and }$p$\textbf{\ are subject to relation }$%
\left( p-1\right) D=4$\textbf{\ (the \textit{Talanov criterion}), and the
norm of the input exceeds a particular finite critical value, }$N_{\mathrm{cr%
}}$\textbf{\ (at larger }$p$\textbf{, the collapse is supercritical, i.e., }$%
N_{\mathrm{cr}}=0$\textbf{). In particular, the critical dimension is }$D=2$%
\textbf{\ for the usual self-focusing cubic nonlinearity (}$p=3$\textbf{),
and }$D=1$\textbf{\ for the quintic case (}$p=5$\textbf{). These NLSEs
create soliton families which are degenerate, in the sense that their norm
takes the single value, exactly equal to }$N_{\mathrm{cr}}$\textbf{.
However, all these solitons are destabilized by the critical collapse.
Accordingly, a problem of great importance is stabilization of solitons in
this class of NLSEs. In particular, a known universal tool for the
stabilization is provided by potential lattices, i.e., spatially periodic
linear potentials. More recently, attention was drawn to \textit{nonlinear
lattices}, which are induced by spatially periodic modulations of the local
nonlinearity strength. However, the use of nonlinear lattices for the
stabilization of solitons turns out to be a challenging issue, as, for
instance, the quintic nonlinear lattice in one dimension provides for
stability of the solitons in a very narrow parameter region. In this work,
we demonstrate that efficient stabilization, not only of spatially symmetric
fundamental solitons, but also of excited states in the form of localized
antisymmetric (dipole) modes, is provided by a combination of the lattice
modulating the quintic term and saturation of the quintic nonlinearity.
Actually, two different species of the dipole solitons are found, which
differ by the location of their center with respect to the nonlinear lattice
(on- and off-site-centered solitons). The most essential result of the
analysis is complete identification of stability areas for all the soliton
species, which is done by means of numerical and analytical methods (the
latter one is based on the well-known \textit{Vakhitov-Kolokolov criterion},
which is a necessary, but not sufficient, condition for the stability of
solitons in self-focusing media). }

\section{Introduction}

Although the possibility of the self-trapping of two- and three-dimensional
(2D and 3D) optical solitons was first predicted half a century ago \cite%
{trap1,trap2,Silberberg}, the creation of such multidimensional localized
modes remains a challenging topic \cite{soliton-bullet,NL,soliton-Rev}. In
particular, a major problem is the fact that the 2D and 3D solitons in media
with the generic Kerr (cubic) nonlinearity are vulnerable to instability
against the critical and supercritical collapse, respectively \cite%
{collapse1,collapse2,collapse3}. As a result, multidimensional solitons are
not routinely observed in experiments, except for spatial \cite{Stegeman}
and spatiotemporal \cite{Wise} quasi-2D beams and pulses in waveguides with
the quadratic nonlinearity, and spatial 2D solitons in bulk media with a
competing nonlinear response of cubic-quintic \cite{Cid-CQ} and
quintic-septimal \cite{Cid-QS} types. Very recently, the creation of 3D
matter-wave solitons in the form of \textquotedblleft quantum droplets",
stabilized by the Lee-Huang-Yang corrections to the mean-field approximation
\cite{LHY}-\cite{Petrov2}, has been reported in Bose-Einstein condensates
(BECs) with long-range dipole-dipole interactions \cite{Kadau}-\cite{Pfau},
as well as in binary condensates with local attraction between the
components \cite{droplets1}-\cite{droplets3}.

Diverse strategies have been theoretically elaborated to stabilize solitons
in 2D and 3D geometries \cite{EPJ-ST}. A versatile stabilization mechanism
relies on the use of periodic (lattice) potentials, which allow one to
predict not only robust multidimensional fundamental solitons, but also ones
with embedded vorticity (alias solitary vortices) \cite{vort1,vort2}.
Available experimental techniques make it possible to readily create
potential lattices in optical media and in BECs, in the form of photonic
crystals \cite{PC} and optical lattices \cite{NL,BEC_OL1,BEC_OL2},
respectively.

Recently, this technique was extended (still, chiefly in the theoretical
form) to the use of nonlinear lattices (NLs), i.e., periodic modulations of
the local nonlinearity strength, as well as their combinations with
linear-lattice potentials \cite{NL}. NLs can be realized in optics and BECs
by means of diverse techniques, e.g., filling voids of photonic crystals
with index-matching materials \cite{NL}, or subjecting BEC to the action of
the Feshbach resonance controlled by a spatially periodic field \cite%
{Feshbach,Feshbach2}.\ In 1D settings, NLs have been used to predict stable
single-hump solitons and various soliton complexes, both single- and
two-component ones \cite{NL1Da, NL1Db, NL1Dc, NL-analy, NL1Dd,vectorNL}. In
the 2D geometry, stable solitons, including fundamental, dipole, and vortex
modes, may only be supported by NLs with sharp edges, such as a period array
of cylinders \cite{circleNL}. Stable solitons supported by NLs with
competing nonlinearities \cite{CQ1D} and combined linear-nonlinear lattices
\cite{LNLa, LNLb, LNLc, LNLd}, have been predicted too.

Similar to the challenging problem of stabilizing 2D solitons in focusing
Kerr media, the occurrence of the critical collapse makes the stabilization
of 1D solitons in self-focusing quintic media a nontrivial problem too \cite%
{quintic, quintic2, quintic3}. This attractive quintic nonlinearity is
physically relevant, as it was predicted \cite{super-TG-theory} and
experimentally demonstrated \cite{super-TG-experiment} that it may be
realized as a super-Tonks-Girardeau gas (a gas of hard-core bosons in a
strongly excited state). On the other hand, the quintic self-focusing in a
nearly \textquotedblleft pure" form can be experimentally realized for the
light propagation in colloidal waveguides with suspensions of metallic
nanoparticles \cite{Cid}.

Periodic lattices acting on the quintic-only self-focusing in 1D cannot
readily stabilize solitons, similar to the above-mentioned situation with
the 2D NLs applied to the cubic self-focusing. As demonstrated in Ref. \cite%
{CQ1D}, in this case the stabilization is possible in a very narrow region
of the solitons' existence area. In this work, we address the existence and
stability of 1D solitons under the action of a NL in a \emph{saturable}
self-focusing quintic medium. Numerical analysis of the model gives rise to
two species of solitons, fundamental and dipole ones (with even and odd
shapes, respectively). In particular, the saturation makes the stability
region of the solitons much broader than it was provided by the NL acting on
the unsaturated quintic term \cite{CQ1D}.

The model is formulated in Section II. The same section also reports some
approximate analytical results for narrow fundamental solitons. Systematic
numerical findings for very broad (unstable) and moderately broad (partly
stable) fundamental solitons, as well as for two types of dipole modes, on-
and off-center-cited ones, are presented in Section III. The paper is
concluded by Section IV.

\section{The model and analytical approximations}

We consider the propagation of light beams in a planar waveguide governed by
the nonlinear Schr\"{o}dinger equation for the field amplitude, $\psi (x,z)$%
:
\begin{equation}
i\psi _{z}=-\frac{1}{2}\partial _{x}^{2}\psi -g\cos (2x)\frac{|\psi
|^{4}\psi }{1+S|\psi |^{4}},  \label{GPE}
\end{equation}%
where $S>0$ determines the saturation of the quintic nonlinearity, and $g>0$
is the strength of the NL, whose period is fixed by scaling to be $\pi $.
Although the model seems quite specific, it may be realized as a solid-phase
sol (similar to ``cranberry glass") with a periodic modulation of parameters
of the dispersed nanoparticles. In this case, the variation of the
parameters, such as the concentration of the nanoparticles and their size,
may change the sign of the nonlinearity \cite{Cid,sol}, which also features
saturation. In fact, a broader class of models similar to the one based on
Eq. (\ref{GPE}) will produce results similar to those reported below. The
nonlinearity is considered here as self-focusing because we consider
solitons with centers placed at $x=0$, around which the sign indeed
corresponds to focusing.

We aim to find stationary solutions of Eq. (\ref{GPE}) with propagation
constant $b$, as $\psi (x,z)=\phi (x)\exp (ibz)$, where real stationary
field $\phi (x)$ is determined by the equation
\begin{equation}
b\phi =\frac{1}{2}\frac{d^{2}\phi }{dx^{2}}+g\cos (2x)\frac{\phi ^{5}}{%
1+S|\phi |^{4}}.  \label{phi}
\end{equation}%
Solitons produced by Eq. (\ref{phi}) are characterized by the total power
(alias norm),%
\begin{equation}
P=\int_{-\infty }^{+\infty }\phi ^{2}(x)dx.  \label{power}
\end{equation}

In the case of small $S$, when the saturation is negligible, the present
model is tantamount to the one considered in Ref. \cite{CQ1D}, where the
above-mentioned narrow stability region for bright solitons was found in a
numerical form. On the other hand, for narrow solitons in the present model,
with width
\begin{equation}
w\ll \pi /4  \label{<<}
\end{equation}
and centered at $x=0$, one may replace $\cos (2x)$ by $1$ in Eq. (\ref{phi}%
). In this case, it is easy to integrate the stationary equation to obtain%
\begin{equation}
b\phi ^{2}=\frac{1}{2}\left( \frac{d\phi }{dx}\right) ^{2}+\frac{g}{S}\left[
\phi ^{2}-\frac{1}{\sqrt{S}}\arctan \left( \sqrt{S}\phi ^{2}\right) \right] .
\label{arctan}
\end{equation}%
It immediately follows from Eq. (\ref{arctan}) that the narrow solitons
exist in the interval of%
\begin{equation}
0<b<b_{\max }\equiv g/S.  \label{bmax}
\end{equation}%
Close to the largest possible propagation constant, \textit{viz}., at $%
0<\delta b\equiv b_{\max }-b\ll b_{\max }$, it follows from Eq. (\ref{bmax})
that the soliton's peak power grows as%
\begin{equation}
\phi ^{2}\left( x=0\right) \approx \left( \pi /2S^{3/2}\right) \left( \delta
b\right) ^{-1},  \label{peak}
\end{equation}%
while its width remains finite, $w\sim 1/\sqrt{2b_{\max }}\equiv \sqrt{%
S/\left( 2g\right) }$, hence the total power (\ref{power}) diverges at $%
\delta b\rightarrow 0$ as
\begin{equation}
P\sim \left( S\sqrt{g}\delta b\right) ^{-1}.  \label{P}
\end{equation}%
Lastly, the above-mentioned condition, $w\ll \pi /4$, which allows one to
replace $\cos \left( 2x\right) $ by $1$, amounts to $S\ll g$.

Under more restrictive conditions, $1\ll \delta b\ll $ $b_{\max }$ (rather
than simply $\delta b/b_{\max }\rightarrow 0$, as considered above), Eq. (%
\ref{phi}) gives rise to an explicit approximate solution in the form of a
compacton \cite{compacton}:%
\begin{equation}
\phi (x)=\sqrt{\frac{\pi g}{4S^{3/2}\delta b}}\left\{
\begin{array}{c}
\cos \left( \sqrt{2\delta b}x\right) ,~\mathrm{at}~~|x|<\pi /\left( 2\sqrt{%
2\delta b}\right) , \\
0,~\mathrm{at}~~|x|>\pi /\left( 2\sqrt{2\delta b}\right) ,%
\end{array}%
\right.
\end{equation}%
with integral power%
\begin{equation}
P=\frac{\pi ^{2}g}{4\left( 2S\delta b\right) ^{3/2}}.  \label{comp}
\end{equation}%
Note that both dependences $P(b)$ given by Eqs. (\ref{P}) and (\ref{comp})
satisfy the Vakhitov-Kolokolov (VK)\ criterion \cite{VK}, $dP/db>0$, which
is a well-known necessary (but, generally speaking, not sufficient)
stability condition for any soliton family supported by self-focusing
nonlinearity.

\section{Numerical results}

\subsection{Fundamental solitons}

In the numerical form, soliton solutions to Eq. (\ref{phi}), both
fundamental (spatially even) and dipole (odd) ones, were produced by means
of the Newton-Raphson iteration method, that provides fast convergence with
a properly selected input. The stability of the solitons was then tested by
means of direct simulations of their perturbed evolution in the framework of
Eq. (\ref{GPE}), using the finite-difference method.

\begin{figure}[tbp]
\begin{center}
\includegraphics[width=1.0\columnwidth]{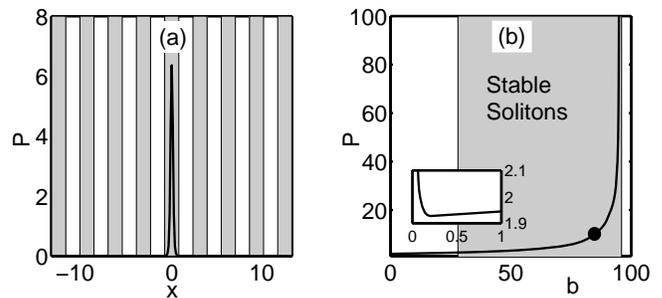}
\end{center}
\caption{(Color online) (a) A typical shape of a stable narrow soliton. (b)
Total power $P$ of the narrow solitons vs. their propagation constant $b$,
for nonlinear strength $g=1$ and saturation $S=0.01$ in Eq. (\protect\ref%
{GPE}). The soliton displayed in (a) corresponds to the bold dot in (b).
Shaded regions in (a), as well as in Figs. \protect\ref{fig2}(a,b) and
\protect\ref{fig5} below correspond to maxima of the local nonlinearity. In
panel (b), as well in Figs. \protect\ref{fig3}, \protect\ref{fig4}, \protect
\ref{fig7}, and \protect\ref{fig8} below, the stability region is shaded.
The inset in (b) and the ones in Figs. \protect\ref{fig3} and \protect\ref%
{fig7} below show curve $P(b)$ at small values of $b$.}
\label{fig1}
\end{figure}

An example of a stable narrow soliton supported by the saturable quintic NL
is shown in Fig. \ref{fig1}(a). The soliton completely localizes itself
within a single cell of the NL, resembling a similar soliton found in the
unsaturated quintic NL model \cite{CQ1D}, i.e., its width $w$ indeed
satisfies condition (\ref{<<}). As predicted above, the stability region of
the narrow solitons, displayed in Fig. \ref{fig1}(b), is close enough to the
largest possible propagation constant $b_{\max }\equiv g/S$, where the
soliton's power may be very large, according to Eq. (\ref{P}). This result
is drastically different from the situation in the unsaturated model, where
the power of stable solitons is restricted to small values \cite{CQ1D}.
Because the NL does not essentially affects the structure and stability of
the narrow solitons, the main objective of this work is, instead, to focus
on the consideration of broader localized states, whose width is comparable
to or larger than the period of the NL.

\begin{figure}[tbp]
\begin{center}
\includegraphics[width=1.0\columnwidth]{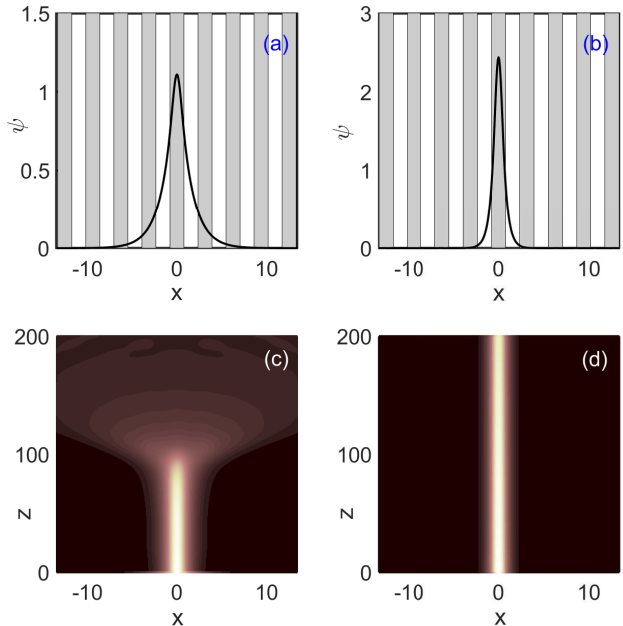}
\end{center}
\caption{(Color online) Typical shapes of unstable (a) and stable (b)
fundamental solitons. Their perturbed evolution is displayed in panels (c)
and (d), respectively. The propagation constant is $b=0.25$ in (a) and $2.5$
in (b), the saturation and NL strength being $S=0.2$ and $g=1$ in both
cases. }
\label{fig2}
\end{figure}

Figures \ref{fig2}(a,c) and (b,d) demonstrate, respectively, that the NL
supports both very broad unstable fundamental solitons, which spread over
several NL periods, and stable moderately broad ones, which are trapped,
essentially, in a single self-focusing stripe. Nevertheless, the latter type
is clearly different from the narrow solitons displayed above in Fig. \ref%
{fig1}. It is relevant to mention that the stability of the solitons was
tested by direct simulations run in a sufficiently large domain, with the
Neumann boundary conditions [see Figs. \ref{fig2}(c,d) and Fig. \ref{fig6}
below]. Although radiation emitted by unstable solitons may partly bounce
back from the boundaries, the size of the integration domain is large enough
to make it sure that the instability of the solitons, if any, is driven by
their internal dynamics, rather than by perturbations induced by reflected
radiation waves.

The difference between the very broad and moderately broad solitons is
natural, taking into account the above-mentioned general fact that NLs may
stabilize solitons under the action of the critical nonlinearity (cubic and
quintic in the 2D and 1D settings, respectively), provided that the
modulation profile is sharp enough: this condition holds for the moderately
broad solitons, while very broad ones feel the action of the underlying
lattice in an averaged, i.e., effectively smooth form \cite{CQ1D}.
\begin{figure}[tbp]
\begin{center}
\includegraphics[width=1.0\columnwidth]{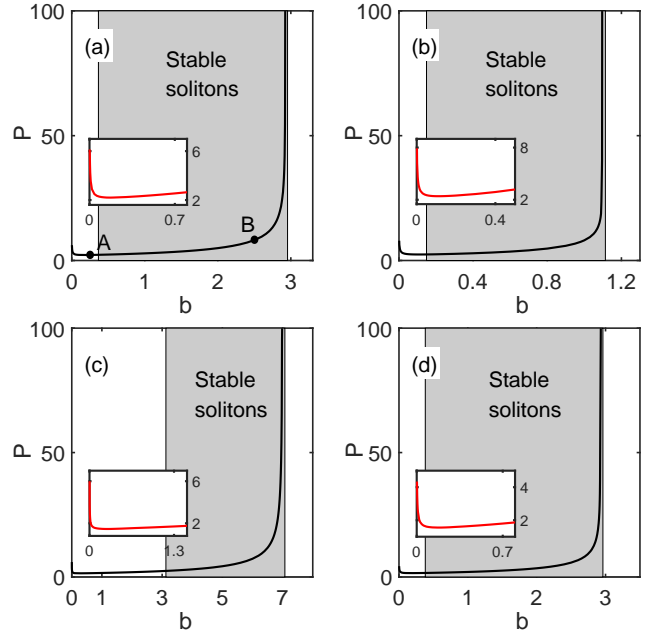}
\end{center}
\caption{(Color online) Soliton power $P$ vs. propagation constant $b$ for
different values of nonlinearity strength $g$ and saturation $S$. (a): $g=1$%
, $S=0.2$; (b): $g=1$, $S=0.4$; (c): $g=2$, $S=0.2$; (d): $g=2$, $S=0.4$.
The solitons are stable in the shaded regions. Points A and B in (a)
correspond to the stable and unstable fundamental solitons displayed in
Figs. \protect\ref{fig2}(a) and (b), respectively.}
\label{fig3}
\end{figure}

The results are summarized in Fig. \ref{fig3}, by dint of plots showing
dependence $P(b)$ for the numerically generated families of the fundamental
solitons. In the same figure, stability regions of the fundamental solitons
are shown too, showing that the stability agrees with the VK criterion, as
stable solitons always obey condition $dP/db>0$. In addition, comparison of
Figs. \ref{fig3}(a) and (b), whose stability regions are, respectively, $%
b\in \lbrack 0.38,2.95]$ and $b\in \lbrack 0.146,1.12]$, shows that the
increase of saturation $S$ leads to decrease of the stability threshold, and
the corresponding cutoff, $b_{\max }$, rapidly decreases too, similarly to
what is predicted for narrow solitons by Eq. (\ref{bmax}). By increasing the
nonlinearity strength $g$, while keeping $S$ fixed, the stability region
broadens, as seen from the comparison of Figs. \ref{fig3}(a,c) and (b, d). A
noteworthy feature of the present model is that, once power $P$ or
propagation constant $b$ become large enough to support stable solitons,
e.g., one reaches the corresponding stability threshold, $b=b_{\min }$
[which is, for instance, $0.38$ in Fig. \ref{fig3}(a)], the fundamental
solitons remain stable up to arbitrarily large values of $P$, i.e., up to
the cutoff propagation constant, $b=b_{\max }$, at which $P$ diverges. This
feature is a natural consequence of the nonlinearity saturation in Eq. (\ref%
{GPE}). The value of $P$ diverges too at $b\rightarrow 0$, as seen from the inside subplot for $P(b)$ relation
in Fig. \ref{fig3} (and the following Fig. \ref{fig6}). The explanation is that, the corresponding soliton's amplitude decreases at $b\rightarrow 0$, the soliton quickly broadens, while the amplitude decreases at a smaller rate.

\begin{figure}[tbp]
\begin{center}
\includegraphics[width=1.0\columnwidth]{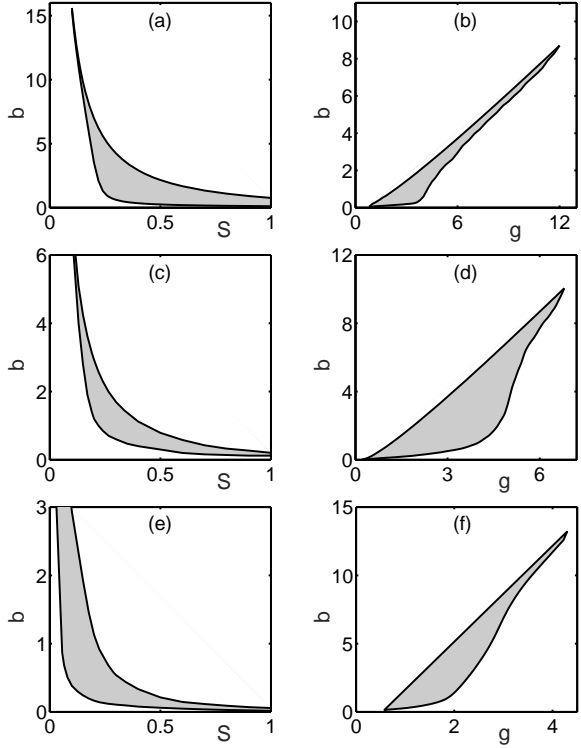}
\end{center}
\caption{(Color online) Stability areas (shaded) for the fundamental
solitons in the $\left( S,b\right) $ plane for different fixed values of the
nonlinearity strength: $g=2$ (a), $g=1$ (c), $g=0.5$ (e), and in the $\left(
g,b\right) $ plane for different values of the saturation parameter: $S=1$
(b), $S=0.5$ (d), $S=0.25$ (f). The fundamental solitons are unstable below
bottom boundaries of the shaded areas, and solitons do not exist above the
top boundaries.}
\label{fig4}
\end{figure}

The results are further collected in Fig. \ref{fig4}, which shows stability
areas for the fundamental solitons in the planes of $\left( S,b\right) $ and
$\left( g,b\right) $. In particular, the trend to the decrease of the
stability threshold, $b_{\min }$, with the increase of $S$, observed in
Figs. \ref{fig4}(a,c,e), is quite natural, as larger values of $S$ make
smaller amplitudes sufficient for affecting the existence and stability of
the solitons through the saturation of the nonlinearity. On the other hand,
the same figures demonstrate that the entire stability areas shrinks with
the increase of $S$ (the same conclusion follows from the comparison of the
stability areas displayed in Figs. \ref{fig4}(b,d,f) for different fixed
values of $S$).

Comparing Figs. \ref{fig4}(b,d,f), one can see that the stability area for
the fundamental solitons shrinks at both very small and very large values of
nonlinearity strength $g$. Further, reducing $S$ leads to an increase of
possible stable region, as can be seen from the planes of $\left( g,b\right)
$ in the right column [Fig. \ref{fig4}(b,d,f)].

\subsection{Dipole solitons}

\begin{figure}[tbh]
\begin{center}
\includegraphics[width=1.0\columnwidth]{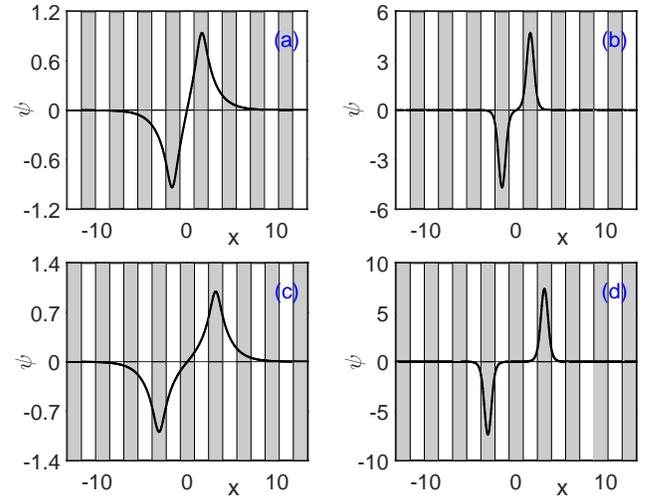}
\end{center}
\caption{(Color online) Examples of dipole solitons, found in the model with
$g=2$ and $S=0.2$: unstable (a) and stable (b) ones with spacing $\protect%
\pi $ between two peaks; unstable (c) and stable (d) ones with spacing $2%
\protect\pi $. The propagation constant is $b=0.3$ in (a) and (c), and $b=6.8
$ in (b) and (d).}
\label{fig5}
\end{figure}

\begin{figure}[tbp]
\begin{center}
\includegraphics[width=1.0\columnwidth]{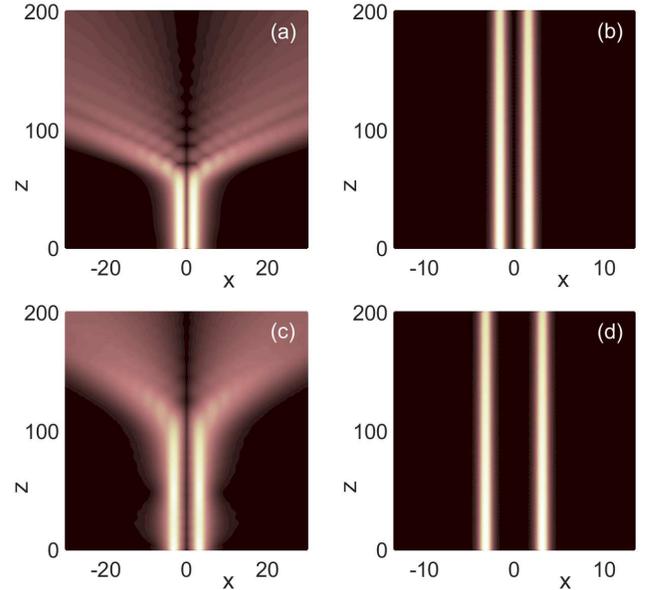}
\end{center}
\caption{(Color online) The perturbed evolution of the dipole solitons whose
stationary shape is displayed in Fig. \protect\ref{fig5}. }
\label{fig6}
\end{figure}

Besides the fundamental solitons, the NL in the present model also supports
two species of localized dipole states, with the spacing between local power
maxima $\pi $ and $2\pi $, which correspond to the single and double lattice
period, severally, see examples in Fig. \ref{fig5} and \ref{fig6}. We call
these soliton modes the off-site-centered and on-site-centered dipole ones,
respectively. By means of the systematic simulations, we have verified that
all dipole modes are unstable (or do not exist) both in the case of the
saturable quintic nonlinearity without the NL [with $\cos (2x)$ replaced by $%
1$ in Eq. (\ref{GPE})], and in the presence of the NL without the saturation
($S=0$), while stabilization can be readily achieved in the present model,
which combines the NL and saturation. Similarly to the fundamental solitons,
the power peaks of unstable dipole modes, of both off- and on-site-centered
types, expand across several cells of the lattice, while the stable ones are
nested in a single well, as seen in Figs. \ref{fig5} and \ref{fig6}.

\begin{figure}[tbp]
\begin{center}
\includegraphics[width=1.0\columnwidth]{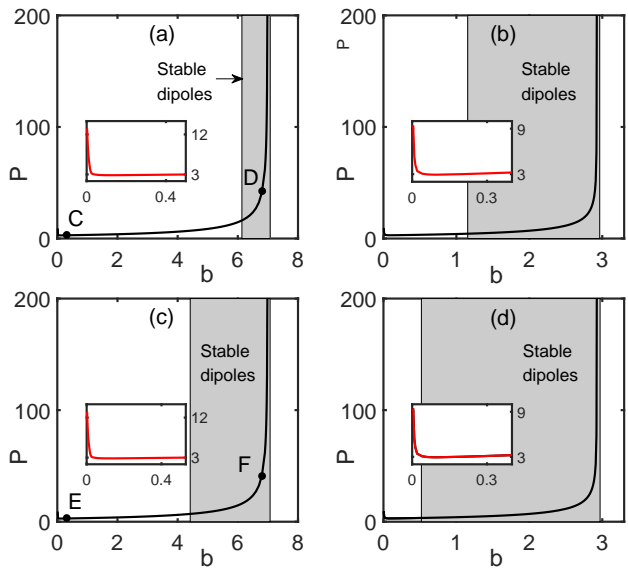}
\end{center}
\caption{(Color online) Power $P$ vs. propagation constant $b$ for the
off-site- (a,b) and on-site- (c,d) centered dipole solitons, respectively.
The saturation coefficient is $S=0.2$ in (a,c), and $S=0.4$ in (b,d), while $%
g=2$ is fixed in all the cases. The dipole solitons are stable in shaded
regions. Points (C,D) in (a) correspond to the dipole modes displayed in
Figs. \protect\ref{fig5}(a) and \protect\ref{fig5}(b), respectively, while
points (E,F) in (c) correspond to Figs. \protect\ref{fig5}(c) and \protect
\ref{fig5}(d).}
\label{fig7}
\end{figure}

\begin{figure}[tbp]
\begin{center}
\includegraphics[width=1.0\columnwidth]{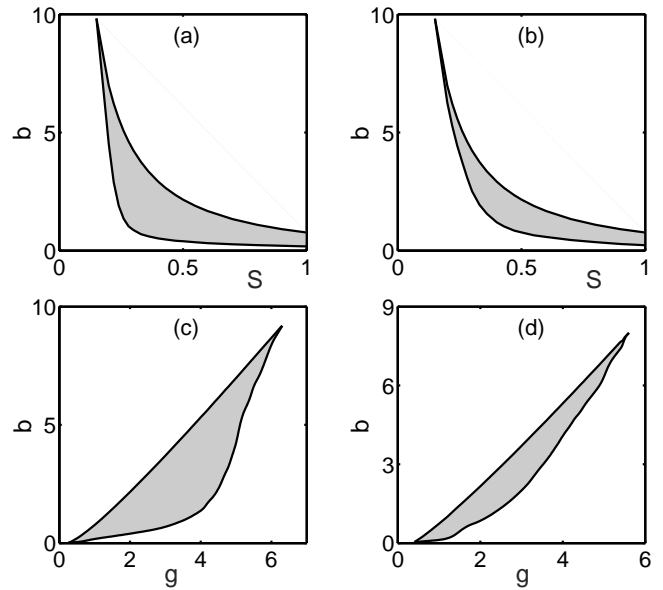}
\end{center}
\caption{(Color online) Stability areas for dipole solitons of the
off-site-centered (a,c) and on-site-centered (b,d) types. Panels (a,b)
display the stability in the $\left( S,b\right) $ plane at fixed $g=2$,
while (c,d) show it in the $\left( g,b\right) $ plane at $S=0.5$.}
\label{fig8}
\end{figure}

Dependences $P(b)$ and respective stability regions for the dipole solitons
are displayed in Fig. \ref{fig7}, and findings for the stability are
summarized in Fig. \ref{fig8}, which displays stability areas in the $\left(
S,b\right) $ and $(g,b)$ parameter planes, cf. similar diagrams for the
fundamental solitons presented above in Fig. \ref{fig3}. The first
noteworthy conclusion is that the cutoff value of the propagation constant, $%
b_{\max }$, above which the stationary solutions do not exist, is precisely
the same for the dipoles of both the off- and on-site-centered types and, in
fact, it is identical to $b_{\max }$ which was found above for the
fundamental solitons, cf. Figs. \ref{fig3} and \ref{fig4}. This conclusion
is explained by the fact that the divergence of $P$ at $b=b_{\max }$ is the
property of the fundamental localized modes, including those of which the
dipoles are built as bound states.

It is seen that, as in the case of the fundamental solitons (cf. Fig. \ref%
{fig4}), the increase of saturation $S$ leads to the decrease of essential
values of $b$, the effect of the variation of $g$ on the stability being
also similar to what was shown above for the fundamental solitons.

On the other hand, Figs. \ref{fig7} and \ref{fig8} clearly show that the
stability threshold (i.e., the bottom boundaries of the stability areas in
Fig. \ref{fig8}) are much lower for the on-site-centered dipole solitons
than for their off-site-centered counterparts, which is explained by the
fact that the modes of the latter type are easier destabilized by the
stronger interaction between the two power peaks, which are separated by a
smaller distance.

Finally, the comparison of Figs. \ref{fig4} and \ref{fig8} demonstrates that
the dependence of the stability areas of the dipoles of both types on the
saturation and nonlinearity coefficients, $S$ and $g$, is qualitatively
similar to that for fundamental solitons. This conclusion is quite natural
too, as the effect of these parameters on the stability is not essentially
different for the solitons of different types.

\section{Conclusion}

In this work, we have investigated a possible way to stabilize three types
of bright solitons, \textit{viz}., the fundamental ones and dipoles with on-
and off-site-centered structure, against the critical quasi-collapse in the
1D medium with the focusing saturable quintic nonlinearity by embedding the
NL (nonlinear lattice) into it. The stabilization is provided by the
interplay between the NL and nonlinearity saturation. For very narrow
fundamental solitons, approximate analytical results were reported, while
the systematic analysis for very broad unstable solitons and moderately
broad partly stable ones was performed by means of numerical methods. In
particular, it was found that the stability region is much broader for the
on-site-centered dipole modes than for their off-site-centered counterparts.
The stability of both the fundamental solitons and both types of the dipoles
conforms to the VK (Vakhitov-Kolokolov) criterion. The predicted
self-trapped states may be realized in nonlinear optical waveguides built of
solid colloidal materials with metallic nanoparticles.

As an extension of the present analysis, it may be interesting to study
possible mobility of solitons in this and similar models.

\section{Acknowledgment}

The work of J. S. and J. Z. was supported by the NSFC, China (project Nos.
61690224, 61690222), by the Youth Innovation Promotion Association of the
Chinese Academy of Sciences (project No. 2016357) and the CAS/SAFEA
International Partnership Program for Creative Research Teams. The work of
B. A. M. was supported, in part, by grant No. 2015616 from the joint program
in physics between the National Science Foundation (US) and Binational
(US-Israel) Science Foundation.

\end{document}